\documentclass[useAMS,usenatbib]{mn2e}

\usepackage{graphicx}

\title[Searching for flickering statistics in T~CrB]
{Searching for flickering statistics in T~CrB}

\author[A. Dobrotka et al.]
{A. Dobrotka$^{1}$\thanks{E-mail: andrej.dobrotka@stuba.sk}, 
L. Hric$^{2}$\thanks{E-mail: hric@ta3.sk}, 
J. Casares$^{3}$\thanks{E-mail: jcv,tsh,igm,tmd@iac.es}, 
T. Shahbaz$^{3}$\footnotemark[3], 
I.G. Mart\'inez-Pais$^{3,4}$\footnotemark[3] and 
\newauthor T. Mu\~noz-Darias$^{3,5}$\footnotemark[3]\\
$^{1}$Departement of Physics, Institute of Materials Science, Faculty of
Materials Science and Technology,\\ Slovak University of Technology in
Bratislava, J\'ana Bottu 25, 91724 Trnava, The Slovak Republic\\
$^{2}$Astronomical Institute of Slovak Academie of Sciences, 05960 Tatransk\'a
Lomnica, The Slovak Republic\\
$^{3}$Instituto de Astrof\'{i}sica de Canarias, La Laguna, Tenerife, Spain\\
$^{4}$Departamento de Astrofísica, Universidad de La Laguna, Tenerife, Spain\\
$^{5}$INAF - Osservatorio Astronomico di Brera, Via E. Bianchi 46, I-23807,
Merate (LC), Italy}

\begin{document}

\date{Accepted ???. Received ???; in original form \today}

\pagerange{\pageref{firstpage}--\pageref{lastpage}}
\pubyear{2008}

\maketitle

\begin{abstract}
We analyze $V$-band photometry of the aperiodic variability in T~CrB. By applying a simple idea of angular momentum transport in the accretion disc, we have developed a method to simulate the statistical distribution of flare durations with the assumption that the aperiodic variability is produced by turbulent elements in the disc. Both cumulative histograms with Kolmogorov-Smirnov tests, and power density spectra are used to compare the observed data and simulations. The input parameters of the model $R_{\rm in}$ and $\alpha$ are correlated on a certain interval and the most probable values are an inner disc radius of  $R_{\rm in} \simeq 4 \times 10^9$\,cm and a viscosity of $\alpha \simeq 0.9$. The disc is then weakly truncated. We find that the majority of turbulent events producing flickering activity are concentrated in the inner parts of the accretion disc.
\end{abstract}


\section{Introduction}

Cataclysmic variables (CVs) are interacting binaries with a white dwarf as the
primary component and a late type main sequence star as the secondary star
which fills its Roche-lobe (see Warner 1995 for review). Consequently mass is
transferred to the white dwarf through the inner Lagrangian point (L$_{\rm 1}$)
and an accretion disc is formed (not in the case of magnetic polars). The
accretion disc is the site where several known physical mechanisms manifest
themselves in the form of short-term stochastic variations in brightness,
generally known as flickering. These oscillations are one of the most
characteristic observational properties of accreting systems with typical
amplitudes ranging from a few dozen of milli-magnitudes up to more than one
magnitude on timescales from seconds to minutes. Symbiotic stars are related
systems to CVs where the secondary star is an evolved red giant where the  mass
loss can be due to either the wind from the giant star or the  Roche-lobe
secondary star, as in "classical" CVs. Thus accretion discs can be formed with
the associated  stochastic variations.

The first systematic study of mechanisms generating the flickering was
performed  by Bruch (1992), where he compared theoretically estimated energies
and timescales for individual mechanisms, with observationally derived values
for a selection of CVs and related symbiotic stars. The author proposed four
possible mechanisms responsible for the observed variations in brightness: (1)
unstable mass transfer from the L$_{\rm 1}$ point and interaction with the disc
edge, (2) dissipation of magnetic loops, (3) turbulences in the accretion disc
and (4)  unstable mass accretion onto the white dwarf. Apparently it was not
enough to restrict the interpretation of the flickering in the whole family of
CVs to only these four mechanisms. For example, in the case of a truncated
accretion disc (intermediate polars) or in the case of the total absence of a
disc (polars) the source of such variations is known to be the accretion stream
threaded by the magnetic field and its subsequent interaction with the surface
of the white dwarf -- accretion shock.

As mentioned earlier, Bruch (1992) performed a systematic study of flickering  activity
in a sample of CVs and symbiotic systems. The promising models which do not
contradict the empirical results, are unstable mass accretion onto the central
accretor or turbulence in the inner accretion disc. Bruch (1996) studied the
variations of photometric data as a function of orbital phase in Z\,Cam. The
source of the flickering was localized in the vicinity of the central accretor,
but the outer disc edge and its interaction with the gas stream from the 
L$_{\rm 1}$ point cannot be neglected (hot-spot). Bruch (2000) studied 4
eclipsing systems: HT\,Cas, V2051\,Oph, UX\,UMa and IP\,Peg.  In these systems,
similar to Z\,Cam, the flickering source  is located in the central part of the
disc or the hot-spot. For V2051\,Oph the region of the flickering was 
confirmed by Baptista and Bortoletto (2004) using eclipse mapping. The hot-spot
is the source of low frequency flickering and the central disc of the high
frequency flickering. Finally, T\,CrB case was studied by Zamanov and Bruch
(1998). The $U$ photometric data showed fast variability located near the
central white dwarf. Following these observational results it is clear that the
dominant source of flickering variability in CVs and T\,CrB is the central part
of the disc or the vicinity of the white dwarf. A promising scenario to explain
this phenomenon is the existence of magneto-hydrodynamic turbulence
transporting the angular momentum outwards (Balbus and Hawley 1998). However
why only the central part of the disc is the source of the observed turbulence?

In this paper we focus on the analysis of turbulent transport of angular
momentum in the disc. We have decided to study T\,CrB due to its rich
phenomenology of rapid variability (i.e. flickering activity) and because of
it's disc properties are suitable for our model.  In Sec.~\ref{observations} we
present our observations. Our treatment of turbulence in the disc as the source
of flickering is explained in Section~\ref{source} and the simulations are
described in Section~\ref{simulations}. Finally the results are analyzed in
Sec.~\ref{analysis} and discussed in Sec.~\ref{discussion}.

\section{Observations}
\label{observations}

The observations of T\,CrB were taken with the 60 cm Cassegrain telescope at
the Skalnat\'e Pleso observatory of the Astronomical institute of Slovak
Academy of Sciences. We used a single channel  photoelectric photometer with
$UBVR$ Johnson filters. We continuously cycled the $UBVR$ filters with 10\,s
integrations per filter and 1\,s dead time  for each filter change, which
results in a time resolution of 45\,s per filter. The data were first used in
the long-term lightcurve analysis by Hric et al. (1998).

\begin{figure}
\includegraphics[width=80mm]{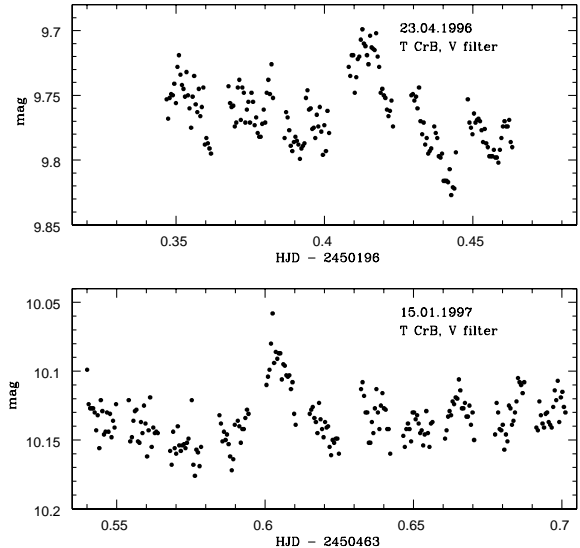}
\caption{Two examples of our $V$-band lightcurves. The upper panel shows data
taken on the 23 April 1998 and the lower panel shows datas taken on 
15 January 1997}
\label{runs}
\end{figure}

Figure~\ref{runs} shows two example runs taken in the Johnson $V$ filter.  The
individual lightcurves consist of rapid aperiodic variability with durations of
$\sim 10^{2}-10^{3}$ seconds. We have in total $\sim 24$ hours of fast
photometry divided into 9 observing runs. Tab.~\ref{obs_log} summarizes all our
observations.

\begin{table}
\caption{The observing log. $T$ is the duration of the observation in hours and
$N$ is the number of data points in the lightcurve.}
\begin{center}
\begin{tabular}{lcccr}
\hline
\hline
 n & Date & HJD-2450000 & $T$ & $N$ \\
 & & (start) & [h] & \\
\hline
 1 & 08.03.1996 & 150.51153 & 1.94 & 120\\
 2 & 09.03.1996 & 151.44714 & 1.80 & 119\\
 3 & 19.04.1996 & 192.35582 & 3.85 & 240\\
 4 & 20.04.1996 & 193.38731 & 3.74 & 240\\
 5 & 22.04.1996 & 195.48539 & 2.30 & 150\\
 6 & 23.04.1996 & 196.34682 & 2.79 & 179\\
 7 & 03.07.1996 & 267.37947 & 1.29 & 89\\
 8 & 11.12.1996 & 428.61449 & 1.80 & 120\\
 9 & 15.01.1997 & 463.54001 & 3.87 & 219\\
\hline
\end{tabular}
\end{center}
\label{obs_log}
\end{table}

\section{Source of aperiodic variability}
\label{source}

T\,CrB is a non magnetic system, therefore the source of the aperiodic
variability must be attributed to one of the four mechanisms proposed by Bruch
(1992). Using integration times of 10-15 minutes,   Zamanov et al. (2005)
discovered variability in the H$_{\rm \alpha}$ emission  line which has a
characteristic timescale of $\sim 1$ hour. Using the velocity information the
authors localize the variability at $\sim 20-30$ R$_{\rm \odot}$. This is
consistent with the disc radius and the gas stream from the $L_1$ point can be
the source of this  variations. Our data also exhibits variability but on
timescales shorter than $\sim 25$ minutes and therefore must have a different
origin.  Furthermore Zamanov and Bruch (1998) and Zamanov et al. (2004) claim
that flickering in T\,CrB with the same timescale as ours arises from the
vicinity of the white dwarf. Therefore, we conclude that unstable mass transfer
from the L$_{\rm 1}$ point is not likely to be the source of our observed fast
aperiodic variability. In our paper we focus on the transport of angular
momentum in the disc. This can be produced by magneto rotational instabilities
(Balbus and Hawley 1998) and we propose that the source of the aperiodic
variability is associated with magnetohydrodynamic turbulence.

\section{Simulations}
\label{simulations}

\subsection{The idea of a turbulent element}

Our idea is based on a simple approximation of mass transfer in the disc. A
turbulent eddy in the disc can be represented by a blob of matter with  mass
$m$, Keplerian tangential velocity at distance $r$
\begin{equation}
v_{\rm t} (r) = r \left( \frac{G M_{\rm 1}}{r^3} \right)^{1/2}
\label{disk_vt}
\end{equation}
and dimension scale $x$. $M_1$ is the white dwarf mass and $G$ is the Newtonian
gravitational constant. The largest possible dimension scale $x$ of such a
turbulent element at radius $r$ from the central object is the scale height of
the disc $H(r)$ at this radius. The blob (which has dimension $x$ and 
a spherical shape approximation) penetrates inwards or outwards 
(a blob with excess of angular momentum moves outwards, whereas a 
blob with deficit moves inwards) with a distance proportional 
to $x$ and a radial viscous velocity $v_{\rm r}(r)$.  The mass of the eddy
$m_{\rm t}$ is therefore the product of the volume and the density $\rho(r)
\simeq \Sigma(r)/H(r)$, where $\Sigma(r)$ is the surface density at distance
$r$ i.e. $\Sigma(r)= \int \rho(r) dz$. Hence,
\begin{equation}
m_{\rm t}(r) = \frac{4}{3} \pi x^3 \rho (r).
\label{hmotnost_turbul}
\end{equation}
Changing the blob's position from $r$ to $r+x$ or $r-x$ also changes it's
Keplerian velocity $v_{\rm t}$ and angular momentum which is defined by
\begin{equation}
{\bf L} = {\bf r} \times m{\bf v_{\rm t}}.
\label{moment_hybnosti}
\end{equation}
Thus by summing up all the turbulent elements we obtain  the global transport
of angular momentum in the accretion disc.
Every radial penetration of a turbulent element produces energy (by
``friction'') which is liberated in the form of an observed flare.  Therefore,
it is expected that the timescale $t$ of the penetration is equal to the
observed flare timescale.

\subsection{Statistical distribution of the eddy dimension scale}

It is assumed that the number of eddies decreases with their dimension scale
$x$. Therefore, we can approximate the distribution function of eddy sizes
$f(x)$ by an exponential function (in analogy to the stochastic movement of
particles in Brownian motion)
\begin{equation}
f(x) = A~{\rm exp}Bx.
\label{exp_funkcia}
\end{equation}
where $B < 0$. The limits of the distribution are $f(0)=1$ and $f(H)=0.01$, where $H$ is the
scale height of the disc. The minimum of the distribution is not 0 because the
exponential function (\ref{exp_funkcia}) cannot reach 0 and thus we obtain

\begin{equation}
f(x) = {\rm exp} \left[ \frac{{\rm ln}~f(H)}{H}~x \right].
\label{exp_funkcia_2}
\end{equation}
Consequently, the distribution of flare timescales $t$ will be given by $f(t)$
with $t \sim x/v_{\rm r}(r)$.

\subsection{Application of the distribution function}

We can define two adjacent rings in a disc, with radius $r$ and $r + \Delta r$
where $\Delta r$ is the ring radial thickness. The ring then has a mass
\begin{equation}
m(r) = 2 \pi r \Delta r \Sigma (r)
\end{equation}
and
\begin{equation}
m(r + \Delta r) = 2 \pi (r + \Delta r) \Delta r \Sigma (r + \Delta r).
\end{equation}
The difference in angular momentum between these two rings is $\Delta L(r) =
L(r + \Delta r) - L(r)$, where the angular momentum of the adjacent rings is
\begin{equation}
L(r) = 2 \pi r^3 \Sigma(r) \Delta r \left( \frac{G M_{\rm 1}}{r^3} \right)^{1/2}
\end{equation}
and
\begin{equation}
L(r + \Delta r) = 2 \pi (r + \Delta r)^3 \Sigma(r + \Delta r) \Delta r \left[\frac{G M_{\rm 1}}{(r + \Delta r)^3} \right]^{1/2}.
\end{equation}
This difference is produced by discrete blobs of diameter $x$ and distribution
$f(x)$, moving stochastically in radial directions between the two adjacent
rings. The angular momentum of an individual blob of size $x$ located at
radius $r$ is then  given by eqs.~\ref{disk_vt} -- \ref{moment_hybnosti}, i.e.
\begin{equation}
L_{\rm t}(r,x) = \frac{4}{3} \pi x^3 \rho (r) r^2 \left( \frac{G M_{\rm 1}}{r^3} \right)^{1/2}
\end{equation}
whereas for a blob at distance $r + \Delta r$, the angular momentum is given by
\begin{equation}
L_{\rm t}(r + \Delta r,x) = \frac{4}{3} \pi x^3 \rho (r + \Delta r) (r + \Delta r)^2 \left[ \frac{G M_{\rm 1}}{(r + \Delta r)^3} \right]^{1/2}.
\end{equation}
Finally the difference in angular momentum is  
$\Delta L_{\rm t}(r,x) = L_{\rm t}(r + \Delta r,x) - L_{\rm t}(r,x)$.

Until now the distribution function has a maximum $f(0) = 1$. This would be
valid for the whole disc only if $\Delta L(r)$ and $H(r)$ were the same (for
all $r$). This is of course not true because of the different disc scale height
$H(r)$ of each ring which determines the largest dimension scale of the
turbulent element. Consequently $f(x)$ is then also a function of $r$, i.e.
$f(r,x)$. The angular momentum difference is different for every pair of rings,
not only because of $H(r)$, but also because of $v_{\rm t}(r)$. Therefore,
$f(r,x)$ must be multiplied by some scale parameter $k = k(r)$ which depends
on radial position of the pair of rings. This parameter is needed for the final
construction of the flare duration histogram; for every $t$ we use $k(r)
f(r,x/v_{\rm r}(r))$ between every pair of rings starting with the inner disc
radius and ending with the outer disc radius (the step $\Delta r$ must be
chosen). Of course, the condition $t \leq H(r)/v_{\rm r}(r)$ must be fulfilled,
i.e. the flare duration cannot be longer than that allowed by the biggest turbulence
dimension scale.

The scaling parameter $k(r)$ can be obtained through the following simple
argument. We can compare $\Delta L(r)$ with the sum of $\Delta L_{\rm t}(r,x)$
between one pair of rings at distance $r$. Summing up angular momentum of all
the eddies is equivalent to integrating $\Delta L_{\rm t}(r,x)$ with the
distribution $f(r,x)$ (equation \ref{exp_funkcia_2}) with the maximum value at
1. We thus obtain
\begin{equation}
\Delta l(r) = \int\limits_{0}^{H(r)} f(r,x) \Delta L_{\rm t} (r,x) dx.
\end{equation}
where this summed angular momentum difference should be equal to the global
angular momentum difference $\Delta L(r)$. The scaling factor $k(r)$ is then 
given by
\begin{equation}
k(r) \Delta l(r) = \Delta L(r)
\end{equation}
which gives
\begin{equation}
k(r) = \frac{\Delta L(r)}{\Delta l(r)}.
\end{equation}
Finally, we obtain the distribution of relative duration by summing up all
durations $t$ for each ring, i.e.
\begin{equation}
\zeta(t) = \sum\limits_{i=1}^{n} k(r_{\rm i}) f(r_{\rm i},t).
\end{equation}
where $n$ is the number of imaginary rings (derived from $R_{\rm in}$, $R_{\rm
out}$ and $\Delta r$) and $r_{\rm i}$ are the ring diameters.

\subsection{The model and free parameters}

For the accretion disc we use the standard Shakura and Sunyaev (1973) model
(see e.g. Frank et al. 1992). The disc scale height, surface density, volume density and radial viscous velocity are then given by the equations
\begin{equation}
H = 1.7~10^8 \alpha^{-1/10} \dot{M}_{\rm acc,16}^{3/20} M_{\rm 1,s}^{-3/8} r_{\rm 10}^{9/8} f^{3/5}~[{\rm cm}]
\label{disk_h}
\end{equation}
\begin{equation}
\Sigma = 5.2 \alpha^{-4/5} \dot{M}_{\rm acc,16}^{7/10} M_{\rm 1,s}^{1/4} r_{\rm 10}^{-3/4} f^{14/5}~[{\rm g~cm^{-2}}]
\label{disk_sigma}
\end{equation}
\begin{equation}
\rho = 3.1~10^{-8} \alpha^{-7/10} \dot{M}_{\rm acc,16}^{11/20} M_{\rm 1,s}^{5/8} r_{\rm 10}^{-15/8} f^{11/5}~[{\rm g~cm^{-3}}]
\label{disk_ro}
\end{equation}
\begin{equation}
v_{\rm r} = 2.7~10^4 \alpha^{4/5} \dot{M}_{\rm acc,16}^{3/10} M_{\rm 1,s}^{-1/4} r_{\rm 10}^{-1/4} f^{-14/5}~[{\rm cm~s^{-1}}]
\label{disk_vr}
\end{equation}
where
\begin{equation}
f = \left[ 1 - \left( \frac{R_{\rm 1}}{r} \right)^{1/2} \right]^{1/4}
\end{equation}
in terms of $r_{\rm 10} = r/(10^{10}\,{\rm cm}), M_{\rm 1,s} = M_1/(1\,{\rm M_{\rm \odot}})$ and $\dot{M}_{\rm acc,16} = \dot{M}_{\rm acc}/(10^{16}\,{\rm g~s^{-1}})$. In our modelling the free parameters are the mass of the central star $M_{\rm 1}$, the viscosity parameter $\alpha$, the inner disc radius $R_{\rm in}$, the outer disc radius $R_{\rm out}$, and the mass transfer rate from the secondary star  which equals the mass accretion rate $\dot{M}_{\rm acc}$ through the disc for a steady state disc. The radius of the white dwarf $R_1$ is
approximated by the  formula given in Nauenberg (1972).

\section{Analysis}
\label{analysis}

\subsection{Input parameters}
\label{input_parameters}

The inner disc radius is a free (searched) parameter and for the starting value
we use the radius of the white dwarf calculated from Nauenberg (1972). The
outer disc radius is first approximated by $\sim 0.5$ of the primary Roche lobe
radius $R_{\rm L1}$ calculated from Paczy\'nski (1971)

\begin{equation}
R_{\rm L1} = 0.462~a~\left( \frac{M_{\rm 1}}{M_{\rm 1} + M_{\rm 2}} \right)^{1/3}~[{\rm cm}],
\end{equation}
where $a$ is the binary orbital separation and $M_1$ and $M_2$ are the primary and
secondary masses respectively. With the orbital period of 227.5 days (Kraft 1958,
Paczy\'nski 1965) this gives a disc radius of up to $\sim 10^{12}$\,cm. Using the
Shakura and Shunyaev (1973) model, for the disc annuli with diameter $10^{12}$\,cm we
obtain an effective temperature of $\sim 1000$\,K and for a disc diameter of
$10^{11}$\,cm of $\sim 5000$\,K (see e.g. Frank et al. 1992).  For effective
temperatures less than 5000\,K (annuli with radius larger than $10^{11}$\,cm) the
contribution to the optical emission is not significant and can be neglected.
Therefore, we can adopt as $R_{\rm out}$=$10^{11}$\,cm  in our analysis for the $V$
filter data.

\begin{figure}
\includegraphics[width=80mm]{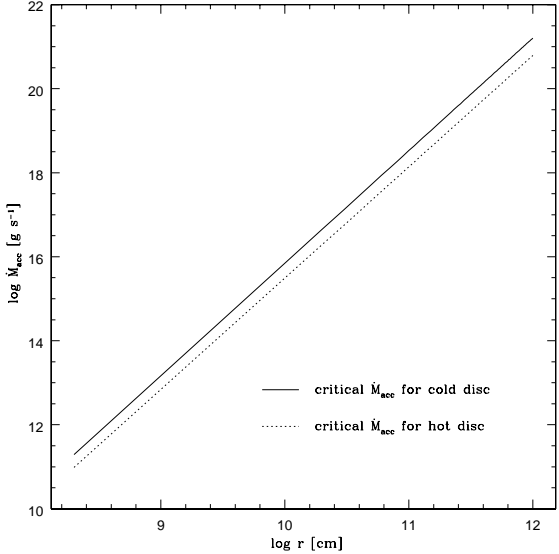}
\caption{
Critical mass accretion rates for the accretion disc in T\,CrB 
as a function of the
distance from the central star $r$.}
\label{dim}
\end{figure}

The white dwarf mass in T\,CrB is $1.37 \pm 0.13$ M$_{\rm \odot}$ (Stanishev et
al. 2004). Using UV observations the authors also derive a mass accretion rate
onto the central accretor of $\dot{M}_{\rm in} = 2.5 \times 10^{17}$~g~s$^{-1}$
during the low-state and $\dot{M}_{\rm in} = 1.9 \times 10^{18}$~g~s$^{-1}$
during the high-state. According to Fig.~4 in Stanishev et al. (2004), all our
observations were taken during high-state and therefore, we assume
$\dot{M}_{\rm in} = 1.9 \times 10^{18}$~g~s$^{-1}$. According to the disk
instability model (see Lasota 2001 for review) this value is higher than the
critical mass accretion rate for the cold disc (Fig.~\ref{dim}) 
for disc radius less than
$10^{11}$\,cm, but lower than the critical mass transfer rate for a hot disc
for larger radius. Stanishev et al. (2004) also proposed that the disc is hot
and optically thick up to a disc radius of $\sim 1$\,R$_{\rm \odot} \sim 7 \times
10^{10}$\,cm. But if there is an intersection of the mass accretion rate with
the critical values, the disc is unstable and should exhibit dwarf novae
activity and cannot be in a steady state. This is not observed and all the
luminosity variations on timescale of days to weeks can be explained by
the activity of the red giant atmosphere (Zamanov and Bruch 1998).

Irradiation by the central hot source can stabilize the disc on the hot branch (see Lasota 2001 for review) to larger disc radii as allowed by the critical mass accretion rate. The white dwarf in the T\,CrB system can have a temperature of about $10^5$\,K (see e.g. Hack et al. 1993) which is very high compared to the typical white dwarf temperature in classical dwarf novae. Hameury et al. (1999) studied the influence of the white dwarf irradiation into dwarf novae cycle. The irradiation temperature is given by
\begin{equation}
T_{\rm irr}^4 = (1-\beta) T_1^4 \frac{1}{\pi} \left[ {\rm arcsin} \rho - \rho (1 - \rho^2 )^{1/2} \right]
\end{equation}
where $\rho = R_1/r$, $T_1$ is the white dwarf temperature and $(1-\beta)$ is the fraction of the incident flux which is absorbed in optically thick regions, thermalized and reemited as photospheric radiation (see e.g. Friedjung 1985, Smak 1989). Taking the absorbed radiation to be of about $\sim 10$\,\% and a white dwarf radius of $0.2 \times 10^9$\,cm we get temperature of 7.700\,K for a distance of $10^{12}$\,cm. This temperature lies in the region of hydrogen ionisation. Therefore permanent hot state is possible up to a radius of $10^{12}$\,cm in the case of $\sim 90$\,\% reflection of incident radiation from the central massive and hot white dwarf.

Finally, Zamanov and Bruch (1998) has claimed that $\dot{M}_{\rm 2} =
\dot{M}_{\rm in}$ in T\,CrB, and the disc can be assumed to be in a hot
steady state. We thus assume that the material is hot, ionized and optically
thick and the conditions for the standard Shakura and Shunyaev (1973) model are
satisfied. Therefore, we adopt $\dot{M}_{\rm acc} = 1.9 \times
10^{18}$~g~s$^{-1}$ for the whole disc.

However, if the disc is irradiated as mentioned earlier, larger disc radius may
have a non-negligible contribution to the $V$-band brightness. We therefore  use
two values ($10^{11}$ and $10^{12}$~cm) for the outer disc radius and study the
differences.

\subsection{Histogram}

\begin{figure}
\includegraphics[width=80mm]{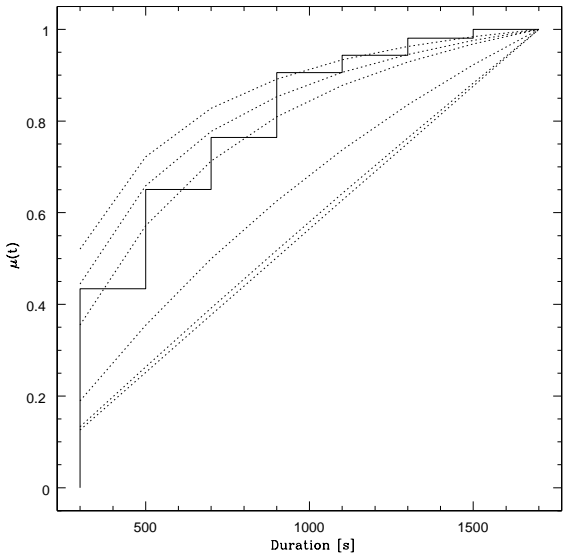}
\caption{
Normalized cumulative histogram of the observed flare duration, together with
6 simulated histograms for different $\alpha$ values; from bottom to top --
0.001, 0.01, 0.1, 0.5, 1 and 10. The simulated histograms have been computed
for the following parameters: $M_1$ = 1.37~M$_{\rm \odot}$, $\dot{M}_{\rm acc}
= 1.9 \times 10^{18}$~g~s$^{-1}$, $R_{\rm out} = 10^{11}$~cm and $R_{\rm in} =
3.5 \times 10^9$~cm.}
\label{hist}
\end{figure}

We have analyzed our data in two different ways, firstly using cummulative 
histograms and secondly using power density spectra. We use the $V$-band data so that
we can compare both methods directly. In this section we compute the cumulative
histogram of the flare duration $t$ from our observed data. We define the flare
duration as the distance between two adjacent minima of a flare. For a flare to be
selected it must be sampled by a minimum number of points. Because we smooth the data
using a 3-point box-car function,  hence this limit must be larger than 3. However,
if we set this limit to be too large, short flares will be missed. After a detailed
investigation, we decided to take a minimum of 3 points either side of the flare
peak. Therefore, the shortest flare will contain 7 points and, because the time
resolution is $\sim$45\,s, they will last for $\sim$315\,s. In Fig.~\ref{hist} we
present a cumulative histogram of the flares, from 300\,s onwards.

In our simulations we need to choose a time step $\Delta t$ to calculate the
final distribution function of duration $t$. Using $\Delta t = 200$\,s and a
minimum flare duration of $t = 300$\,s we obtain a synthetic cumulative histogram
which can be compared with the observed data using a Kolmogorov-Smirnov test (KS
test). This test searches for the maximal difference $D$ between two cumulative
histograms;

\begin{equation}
D = {\rm max} \left| O_{\rm i}(t) - S_{\rm i}(t) \right|
\end{equation}
where $O_{\rm i}(t)$ is the observed histogram and $S_{\rm i}(t)$ is the
simulated distribution function/histogram.

\begin{figure}
\includegraphics[width=55mm,angle=-90]{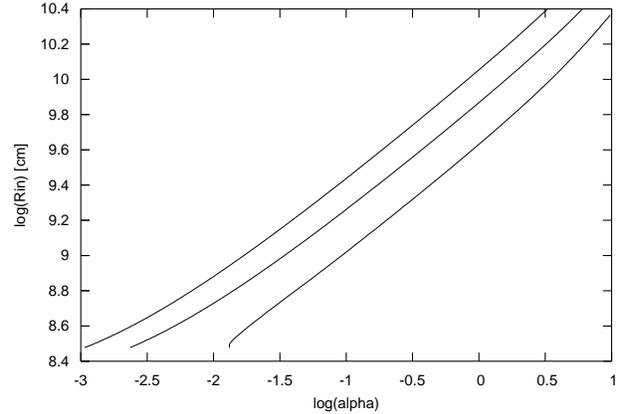}
\caption{Contour plots of KS test for KS values 0.3, 0.2 and 0.1 (from left to right).}
\label{isocont_ks}
\end{figure}

Some examples of simulated histograms are shown in Fig.~\ref{hist}. The
simulations were computed for the parameter $\alpha$ in the interval 0.001 --
10. Fig.~\ref{isocont_ks} shows  the results of the KS test in the 
$\alpha$--$R_{\rm in}$ plane.It is clear that
this method is not sensitive enough to constrain the two free parameters. The
best KS test value 0.1 is too wide and so there are no closed contours. 
Reducing the KS value on the Fig.~\ref{hist} down to $\sim 0.0001$ does 
not show more contours features and so
this method is clearly is not suitable for our study.

\subsection{Power Density Spectrum}

\begin{figure}
\includegraphics[width=80mm]{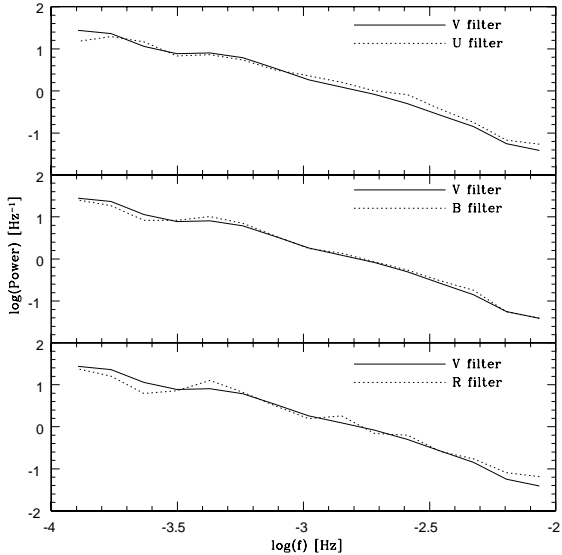}
\caption{The power density spectrum of our observed data in different 
Johnson filters compared to the $V$-band data. }
\label{pds_filters}
\end{figure}

Our method of flare identification can be very subjective because different
timescales are superimposed. Therefore we decided to analyze our data in a
another way by calculating the power density spectra (PDS) of our  observations
and the synthetic data. We applied the Lomb-Scargle algorithm (Scargle 1982)
which is adequate to deal with unevenly spaced data. Fig.~\ref{pds_filters} show
PDS of our data in $UVBR$ Johnson filters. The PDS of the $B$ and $V$-band
lightcurves are very similar. Whereas the PDS for the other filters show some
fluctuations which can be caused by the larger scatter/noise in $U$ color and the
lower amplitude of the variability in the $R$-band. Therefore, we decided to
focus on the $V$-band data, which is also adequate after the outer disc estimate
in Sec.~\ref{input_parameters}.

Fig.~\ref{pds_rout} shows examples of PDS for different viscosity and outer disc
radius. The value for the outer disc radius of interest are $10^{11}$ and
$10^{12}$\,cm. Differences in $R_{\rm out}$ only start to be visible for high
value of $\alpha$.  The two $R_{\rm out}$ cases are indistinguishable for $\alpha
= 0.5$ and 2.5 (used as upper limit of viscosity in the following PDS study).
Only for $\alpha = 5.0$ can the two radii be distinguished. We therefore adopt
this viscosity value as our sensitivity limit. for the outer disc radius in our
interval of interest. According to the disc instability model (see e.g. Lasota
2001 for review) accretion discs typically have $\alpha < 1.0$, hence our
interval of interest is lower than the sensitivity limit. Therefore, this
parameter has no influence in our simulation, and so we fixed it to $R_{\rm out}
= 10^{11}$\,cm.

\begin{figure}
\includegraphics[width=80mm]{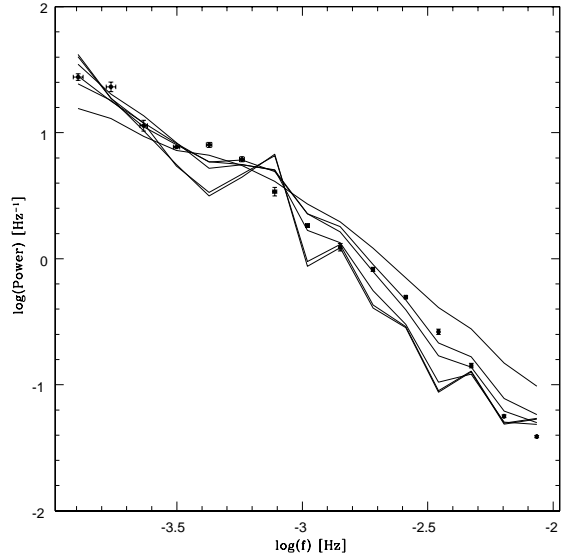}
\caption{The mean power density spectrum of the observed runs 
(the solid points with error bars) compared with 6 examples of 
simulated power density spectra (solid line) for 
$\alpha = 0.001, 0.01, 0.1, 0.5, 1$ and 10 (from bottom to top). 
}
\label{pds_examples}
\end{figure}

\begin{figure}
\includegraphics[width=80mm]{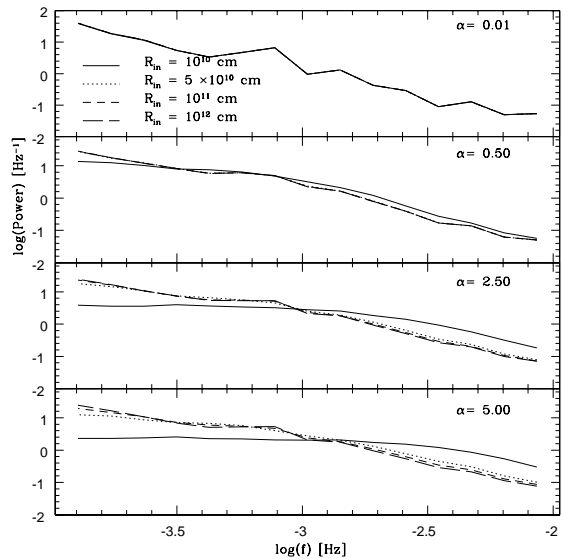}
\caption{
Examples of simulated power density spectra for different viscosity 
and outer disc radii.
The inner disc radius has been fixed to $R_{\rm in} = 3.5 \times 10^9$~cm.}
\label{pds_rout}
\end{figure}

\begin{figure}
\includegraphics[width=55mm,angle=-90]{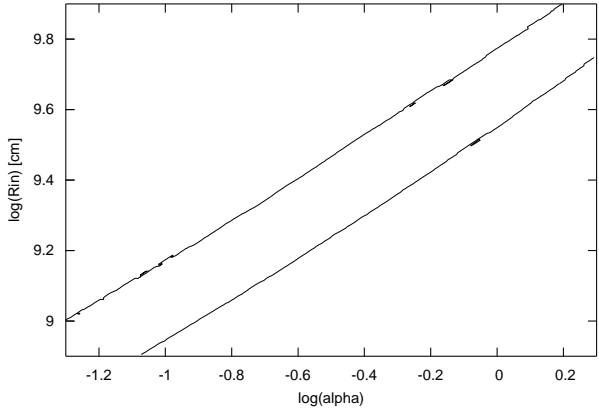}
\caption{The $\chi^2$ contours of the power density spectra 
at the 90\% confidence (1.6-$\sigma$) level.}
\label{isocont_pds}
\end{figure}

For a given pair of the input parameters $R_{\rm in}$ and $\alpha$ we produced
100 synthetic runs with the flare duration statistics calculated using the method 
described in Sec.~\ref{simulations}. The flares has a triangular shape and the time
sampling of the runs were sampled the same as the observed runs (a few examples are
depicted in Fig.~\ref{synt_runs}). We then calculated a mean PDS.
Fig.~\ref{pds_examples} shows six examples of the simulated PDS for $\alpha = 0.001 -
10$. We subsequently calculated the difference in $\chi^2$ between the observed and
simulated PDS as follows
\begin{equation}
\chi^2 = \sum\limits_{{\rm i}=1}^{\rm N}\frac{(o_{\rm i} - c_{\rm i})^2}{e_{\rm i}},
\end{equation}
where $o_{\rm i}$ and $c_{\rm i}$ are the observed and calculated values
respectively, $e_{\rm i}$ is the error of the observed PDS value and N is the
number of bins in the PDS (we used 15). We then computed a 2-D $\chi^2$ grid 
fits of the
inner disc radius and viscosity parameter (Fig.~\ref{isocont_pds}) 
for the same system parameters as for the KS test. This
method is clearly more suitable for our study as it is more sensitive to the
change in input parameters and the wide plateau seen the KS test is not
present. On the contrary the contour-features are well localized and show
that the input parameters $\alpha$ and $R_{\rm in}$ are correlated at some interval.


A minimum $\chi^2$ of 282 is found at $R_{\rm
in} \simeq 3.9 \times 10^9$\,cm and $\alpha \simeq 0.9$.
Fig.~\ref{isocont_pds} shows the 90\% probability (1.6-$\sigma$) confidence levels
in the $\alpha - R_{\rm in}$ plane after rescaling so that the reduced $\chi^2$ is 1. The parameters show a linear
trend, an indication that the two are correlated. The correlation is constrained to
the interval $R_{\rm in} \simeq 1.1 - 4.3 \times 10^9$\,cm for all acceptables 
$\alpha$ values for a disc in hot branch, i.e. 0.1 -- 1.0.

\section{Discussion}
\label{discussion}

The 90\% confidence level contours of the PDS fits demonstrates that the inner disc
radius and the viscosity parameter are correlated in the interval $R_{\rm in} \simeq
1.1 - 4.3 \times 10^9$\,cm for all aceptable $\alpha$ values for a disc in hot 
branch, i.e. 0.1 -- 1.0. Our best solution $R_{\rm
in} \simeq 3.9 \times 10^9$\,cm and $\alpha \simeq 0.9$ is in agreement with the
inner disc radius obtained by Skopal (2005) through modeling of UV/optical/IR spectra
of T\,CrB; i.e. $0.05 - 0.10$\,R$_{\rm \odot} \simeq 3.5 - 7.0 \times 10^9$~cm. This
is larger than the estimated white dwarf radius and hence the disc seems to be
truncated. Furthermore, the best model viscosity parameter is in agreement with 
expectations for a disc in the permanent hot branch,
i.e. $0.1 < \alpha < 1.0$.

Our best values are derived from the minimal value of $\chi^2$. The 90\%
contours levels suggest a large uncertainty in parameters which are correlated. The
determination of one of the two parameters requires an independent measurement of the
other. Taking the inner disc radius from Skopal (2005) to lie in the interval of
$3.5 < R_{\rm in} < 4.3 \times 10^9$\,cm then gives a viscosity interval of $0.7 <
\alpha < 1.0$.
King et al. (2007) report, for fully ionized geometrically thin
accretion discs, that the observational evidence suggests $\alpha \sim 0.1 - 0.4$. In
particular, the simulations for SS\,Cyg (Schreiber et al. 2003) yield a viscosity
value of about $0.1 - 0.2$, whereas Schreiber and G\"ansicke (2002) obtain $\alpha$ =
0.5 for SS\,Cyg in the ionized hot state. Our derived viscosity value of $0.7 - 1.0$
is considerably higher than previous values. However, Schreiber et al. (2003) used a
maximum mass transfer rate of $\sim 1 \times 10^{17}$~g~s$^{-1}$ in his simulations.
This is one order of magnitude lower than mass accretion rate in T\,CrB. The high
mass transfer rate through the disc together with the irradiation by the central
source may explain the higher temperature and viscosity than in standard dwarf nova
system. Furthermore the lower viscosity values were determined from the long-term light
curve behaviour where the considered timescales are typical for the largest disc
radii, whereas our constraints apply mainly at the inner disc radius. This gives
an idea that the viscosity parameter $\alpha$ can also be a function of the
distance $r$. In the standard disc instability model (see e.g. Lasota 2001 for
review) the main difference between the viscous parameter in the cold and hot
branch is due to the ionisation of hydrogen. Once the material is in the hot
state, the ionization state of hydrogen does not change with increasing
temperature and the viscosity parameter is supposed to be constant in the whole
disc. Intensive studies of the disc behaviour with constant  $\alpha$ were
performed in the case of dwarf novae and low mass X-ray binaries.  But the typical
disc radii of these binaries are of the order of $10^{10} - 10^{11}$\,cm. The case
of symbiotic systems, such as T\,CrB is different  because of the  large binary
separation and the possible outer disc radii up to $10^{12}$\,cm. But we showed
that our simulations are not sensible to such large disc radii in the acceptable
$\alpha$ regime. The outer disc radius used for T\,CrB is $10^{11}$\,cm, and hence
is in agreement with the radius of low mass X-ray binaries modelled by Lasota
(2001) with constant $\alpha$ values. Therefore we conclude that radial variations
of the viscous parameter is not required. An alternative explanation for the
different $\alpha$ results could be a different model other than the Shakura and
Shunyaev (1973) model for the large outer discs in symbiotic systems.
An application of our simulations to dwarf novae with disc radii of
$10^{10}$\,cm can test whether the $\alpha$ differences remain or disappear. This
case is problematic because dwarf novae discs are not in a steady-state.
Another possibility is to study dwarf novae in outburst when the disc is fully
ionised and is in a quasi steady-state (see e.g. Lasota 2001) and in a
permanent hot state except for in the lightcurve minima e.g. in VY\,Scl systems (see e.g.
Warner 1995).

\begin{figure}
\includegraphics[width=80mm]{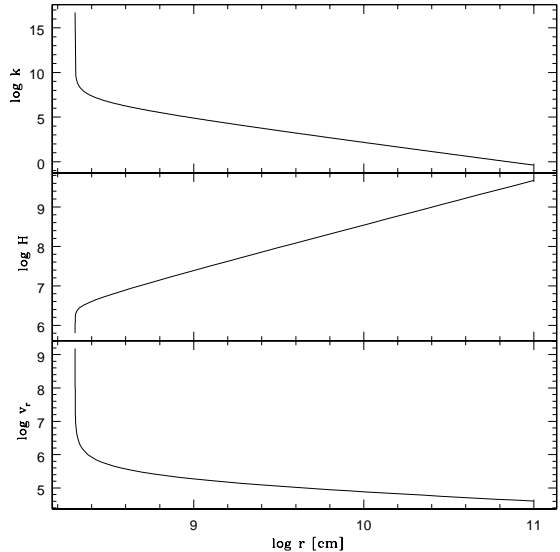}
\caption{A simulation of the scaling parameter $k$, disc scale height $H$ and radial
velocity $v_{\rm r}$ for $M_1$ = 1.37~M$_{\rm \odot}$, $\dot{M}_{\rm acc} = 1.9
\times 10^{18}$~g~s$^{-1}$ and $\alpha = 0.5$.}
\label{disc_prof}
\end{figure}

Another interesting question is the location of the flickering events in the inner
disc. This can be understood from the behavior of the scaling parameter $k(r)$
(Fig.~\ref{disc_prof}). For very small $r$ values, in the vicinity of the central
star, $k(r)$ rises very steeply because of the small disc scale height
(Fig.~\ref{disc_prof}) and hence small turbulent eddies. Furthermore, $k(r)$
decreases with increasing $r$ and the disc scale height, and thus less events are
needed to transport the angular momentum. This strongly suggests that the inner parts
of the disc are the source of the fast flickering, which is in agreement with
observations (see e.g. Bruch 1992, Zamanov \& Bruch 1998, Bruch 1996, Bruch 2000,
Baptista \& Bortoletto 2004, Zamanov 2004). A similar behavior is deduced from the
sensitivity of our modeling to the input parameters $R_{\rm in}$ and $R_{\rm out}$,
which is very high for $R_{\rm in} \sim 10^9 - 10^{10}$\,cm but not for $R_{\rm out}
\sim 10^{11} - 10^{12}$\,cm. This suggests that the majority of turbulent events are
located in the central part of the disc up to radial distance $\sim 10^{10}$\,cm.

The inner disc radius and the viscosity is correlated only for low values of $r$
comparing to the disc radius. This can be explained because the viscosity must
increase with higher $R_{\rm in}$ in order to get the same $\chi^2$ residual. This
means that by truncating the central part of the disc, we are removing many fast
events which should be compensated by the increase of the radial velocity of slower
events at larger radii. The radial velocity $v_{\rm r} \sim \alpha^{4/5}$ strongly
decreases at very small $r$ (see Fig.~\ref{disc_prof}). An alternative explanation
relies on the total number of events in the central disc. The scaling parameter $k$
gives an estimate of the number of events and this is very large at small $r$. Hence
by truncating the central part of the disc we are removing many events. Whereas at
larger $r$ the decrease of $k$ and $v_{\rm r}$ is flatter and the correlation between
$R_{\rm in}$ and $\alpha$ is not as strong as for small values of $r$.

\begin{figure}
\includegraphics[width=80mm]{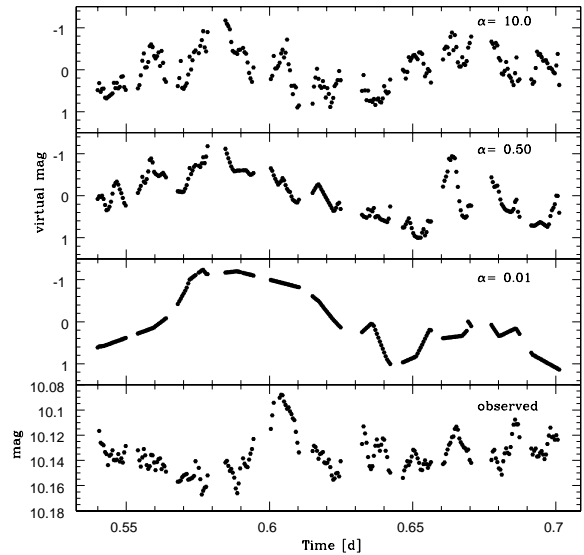}
\caption{
Three examples of the synthetic data compared to an observed lightcurve. 
The simulated
runs are computed for $M_1$ = 1.37~M$_{\rm \odot}$, $\dot{M}_{\rm acc} = 1.9 \times
10^{18}$~g~s$^{-1}$ and $R_{\rm in} = 3.5 \times 10^9$~cm and three different
viscosities $\alpha = 0.01$, 0.50 and 10.0.}
\label{synt_runs}
\end{figure}

Fig.~\ref{synt_runs} shows three examples of simulated lightcurves computed for
different viscosities. It is obvious that high frequency fluctuations increase with
$\alpha$, whereas low-frequency fluctuations decrease with $\alpha$. This
disappearance of large events is visible in our PDS (Fig.~\ref{pds_examples}). The
low viscosity PDS has a strong peak near frequency $10^{-3}$~Hz which is caused by
the quasi-periodic oscillations present in the simulated data. Such variations has a
timescale of $\sim 0.01$ day clearly visible on Fig.~\ref{synt_runs} for $\alpha =
0.01$ case. The timescale of the data is too short to distinguish whether the
fluctuations are stochastic or quasi-periodic. For higher viscosities,
large events disappear and so does the evidence for quasi-periodic signals in the
simulated data. Hence the peak at $10^{-3}$~Hz dissapear.

In addition, the observed PDS in Fig.~\ref{pds_examples} are scattered which produce a high $\chi^2$. This
behavior tells us something about the observational demands. Our observed PDS is
computed from 9 runs, whereas the simulated PDS is computed from about 100 synthetic
runs. Smoother observed PDS (made from a higher number of observational data sets)
are required for the analysis to get lower $\chi^2$ and then make the method more sensitive to the input parameters. The mean
duration of our data is $\sim 2.6$ hours. Longer observations could remove the
quasi-periodic signals at low viscosities and reduce the scatter in the observed and
simulated PDS at low frequencies ($\sim 10^{-4}$~Hz). Using of a single filter can
improve the data resolution to $\sim 10$ seconds and hence the quality of the PDS at
higher frequencies ($\sim 10^{-2}$~Hz). Also the observational gaps needed to observe
a comparison star (clearly visible on Figs.~\ref{runs} and \ref{synt_runs}) affect
the PDS. Therefore CCD observations with fast readout speeds are preferred.

Our best model yields a parameters in reasonable agreement with other works. We
therefore conclude that our simple idea of angular momentum transport through
turbulent eddies is sufficient to explain the observed flickering statistics. No
other mechanisms are then required to generate the observed flickering activity.

\section{Summary}

We have analyzed $V$-band photometry of the aperiodic variability of T\,CrB during
the high state, taken using data with a time resolution of $\sim 45$\,s. We
developed a method to simulate flickering activity based on a simple idea of angular
momentum transport through turbulent processes in the accretion disc. We simulated
artificial data sets and compared them with the  observations using cumulative
histogram with Kolmogorov-Smirnov tests and power density spectra. The best model
yielded $R_{\rm in} \simeq 4 \times 10^9$~cm and $\alpha \simeq 0.9$ but with large uncertainties. However,
using the results from literature for the inner disck radius, we derived the 
viscosity to lie in the interval $0.7 < \alpha < 1.0$. The disc is weaklu truncated 
and we provide an explanation for the concentration of flickering events in
the central part of the accretion disc. The correlation between the parameters $R_{\rm in}$ and $\alpha$ 
is explained by
the coupling between the number of fast events and the radial velocity, i.e. the
removal of fast events produced by increasing $R_{\rm in}$ is compensated by an
increase in $v_{\rm r}$ and hence $\alpha$. Our turbulent scenario of angular
momentum transport successfully explains the observed flickering statistics without
any other mechanism required.

\section*{Acknowledgment}

AD and LH acknowledge the Slovak Academy of Sciences Grant No. 2/7011/7. AD also
acknowledge HPC Europa grants HPC04MXW87 and HPC0477ZZL for supercomputing training
and V.Antonuccio-Delogu from INAF Catania for computer providing. JC, TS, IM and TMD
acknowledge support from the Spanish Ministry of Science and Technology  under the
grant AYA\,2007\,66887. 
Partially funded by the Spanish MEC under the Consolider-Ingenio 2010  Program grant 
CSD2006-00070: ``First Science with the GTC'' 
(http://www.iac.es/consolider-ingenio-gtc/).

\end{document}